\documentclass[%
mathleft,%
final,%
]{an}
\usepackage{graphicx,hyperref}
\usepackage{times}
\begin{document}

\Pagespan{1}{}
\Yearpublication{2013}%
\Yearsubmission{2012}%
\Month{1}%
\Volume{334}%
\Issue{1}%
\DOI{This.is/not.aDOI}%

\title{The SLoWPoKES Catalog of Low-mass Ultra-wide Binaries:\\ 
  A Cool Stars Resource for Testing Fundamental Properties and for Constraining Binary Formation Theory}

\author{Saurav Dhital \inst{1,2}\fnmsep\thanks{Corresponding author. 
  \email{saurav.dhital@vanderbilt.edu}}
\and  Andrew A. West\inst{1}
\and Keivan G. Stassun \inst{2,3}
\and Nicholas M. Law \inst{4}
}
\titlerunning{The SLoWPoKES Low-mass Ultra-wide Binaries}
\authorrunning{Dhital et al.}
\institute{
Department of Astronomy, Boston University, 725 Commonwealth Avenue, Boston, MA 02215, USA
\and
Department of Physics \& Astronomy, Vanderbilt University, 6301 Stevenson Center, Nashville, TN, 37235, USA
\and
Department of Physics, Fisk University, 1000 17th Avenue N., Nashville, TN 37208, USA
\and 
Dunlap Institute, University of Toronto, 50 St. George St, Toronto, ON, Canada}

\received{XXXX}
\accepted{XXXX}
\publonline{XXXX}


\abstract{%
We present results from the Sloan Low-mass Wide Pairs of Kinematically
Equivalent Stars (SLoWPoKES) catalog of ultra-wide (10$^{3-5.5}$~AU),
low-mass (K5--M7) common proper motion binaries.  We constructed a
Galactic model, based on empirical stellar 
number density and 3D velocity distributions, to select bona fide
pairs with probability of chance alignment $<$5\%, making SLoWPoKES an
efficient sample for followup observations. Our initial
catalog contains 1342 disk dwarf, subdwarf, and white dwarf--red dwarf
systems and is the largest collection of low-mass, wide binaries ever
assembled. The diversity---in mass,
metallicity, age, and evolutionary states---of SLoWPoKES pairs makes
it a valuable resource of {\it coeval laboratories} to examine and
constrain the physical properties of low-mass stars. SLoWPoKES
pairs show signatures of two (or more) formation modes in the
distribution of the physical separation and higher-order
multiplicity. Neither dynamical dissipation of primordial
triples/quadruples or dynamical capture of ejected stars can explain
the observed populations by itself.
We use followup spectroscopic observations to recalibrate the
metallicity-sensitive $\zeta_{\rm{TiO/CaH}}$ index by assuming that both
members of the binary system have the same composition. Our
new formulation is a significantly better tracer of absolute metallicity,
particularly for the early-type M dwarfs. The catalogs are publicly
available on a custom data visualization portal.}

\maketitle

\section{Introduction}
Often times it is the extremes of a distribution that encode the most
valuable information about the underlying physics. While typical
binary star separations are $\sim$30~AU (e.g., Duquennoy \&
Mayor~\cite{Duquennoy1991}; Fischer \& Marcy~\cite{Fischer1992}), a large 
number of ultra-wide binary systems ($\geq 10^{3-5}$~AU) have
been identified in star-forming regions
(e.g., Connelley, Reipurth \& Tokunaga~\cite{Connelley2008}; Kraus \&
Hillenbrand~\cite{Kraus2009b}) and in the field
(e.g., Chaname \& Gould \cite{Chaname2004}; L{\'e}pine \&
Bongiorno~\cite{Lepine2007a}; Dhital et al.~\cite{Dhital2010}).  While these
ultra-wide systems make up $\leq$10\% of all binaries, they have
uniquely interesting and valuable roles as boundary conditions to the star
formation process. From detailed numerical simulations, we now know star
formation to be a dynamic process where cores and protostars interact with each other
and modify the environments (e.g., Bate~\cite{Bate2009}). Therefore,
how these wide binaries are formed has important implications on our
understanding of the entire star formation process as well as that of the 
sub-structure in star-forming regions and how the field gets populated
with single and binary stars. In addition, whether planets can be
formed and/or can survive in such dynamically active environments is
an important question.
Wide pairs are important coeval laboratories, and understanding
their origins informs us about the validity of their coevality
(e.g., Stassun, Mathieu \& Valenti~\cite{Stassun2007}; Kraus \&
Hillenbrand~\cite{Kraus2009a}) and iso-metallicity (e.g., L{\'e}pine, Rich
\& Shara~\cite{Lepine2007b}; Dhital et al.~\cite{Dhital2012}) assumptions.  

We have assembled the Sloan Low-mass Wide Pairs of Kinematically Equivalent
Stars (SLoWPoKES) catalog of wide, low-mass common proper motion
binary systems; the selection is briefly described in Section
\ref{Sec:slowpokes}. We have identified signatures of two (or more)
formation modes in the physical separation and higher-order
multiplicity distributions of SLoWPoKES systems, which we describe in
Section \ref{Sec:formation}. In Section \ref{Sec:metallicity} we
present followup spectroscopic observations of SLoWPoKES systems to
verify and refine the low-mass metallicity index,
$\zeta_{\rm{TiO/CaH}}$. We have created a web portal for data
visualization and dissemination, which we describe in Section
\ref{Sec:portal}. We provide a synopsis of our ongoing  and future
work in Section \ref{Sec:future}.

\section{Construction of the SLoWPoKES Catalog}\label{Sec:slowpokes}
We searched for common proper motion companions around
low-mass dwarfs (spectral type of K5 or later) in
the Sloan Digital SKy Server (SDSS) Data Release 7 photometric catalog 
(Abazajian et al.~\cite{Abazajian2009}). By matching photometric distances
(Bochanski et al.~\cite{Bochanski2010}) and SDSS/USNO-B proper motions
(Munn et al.~\cite{Munn2004}),
we identified candidate systems with angular separations
of 7--180$\arcsec$. A series of quality cuts on the photometry and the
proper motions were made to ensure the resultant sample was not
contaminated. See Dhital et al.~(\cite{Dhital2010}) for details.

False positives are inherent in such statistical samples, arising from
chance alignments within the uncertainties of the selection
criteria. The probability of such a chance alignments increases with the
separation of the candidate companion as well as with the distance to
the star. Therefore, for a pure sample of physically associated CPM
pairs to be constructed, a detailed analysis of the probability of
chance alignment need to be conducted. We constructed an Monte
Carlo-based Galactic model using empirical stellar number density
(Juric et al.~\cite{Juric2008}; Bochanski et al.~\cite{Bochanski2010})
and 3D velocity distributions (Bochanski et
al.~\cite{Bochanski2007a}). Each iteration of the model repopulates 
a $30\arcmin\times30\arcmin$ conical volume extending to 2500~pc and
centered on the primary of each candidate system and assigns a 3D
velocity to each star. As only single stars are included in this model
galaxy, any star with matching position and proper motions to the
primary of the candidate system is a chance alignment. The probability
of chance alignment (P$_{\rm f}$) is then, simply, how many times such matches were
found in $10^5$ iterations. Only pairs with P$_{\rm f} \leq$5\% were
included in the SLoWPoKES catalog. Thus, SLoWPoKES contains only bona fide
binary systems, making it an ideal source for followup observations. A total
of 1342 common proper motion binaries, comprising of GK--M dwarf, M
dwarf, M subdwarf, and white dwarf--M dwarf systems, were
identified. They span a wide range in mass, mass ratio, metallicity,
and evolutionary states.

Radial velocities from followup spectroscopic observations have
confirmed the physical association of observed SLoWPoKES pairs
(Dhital et al.~\cite{Dhital2012}). 

\section{Multiple Modes for Formation of Wide Binaries}\label{Sec:formation}
\begin{figure}
  \includegraphics[angle=0, width=\linewidth]{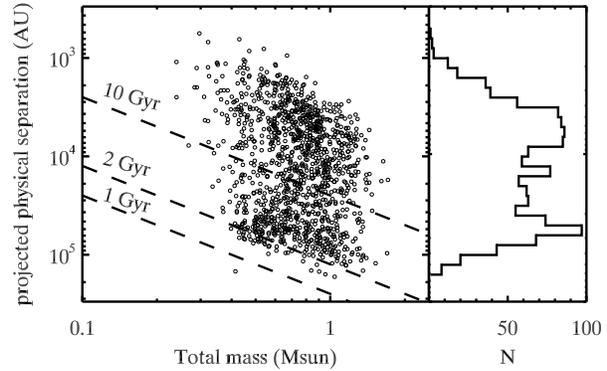}
  \caption{Projected physical separation vs. total mass (inferred from
    $r-z$ colors) for the M dwarf pairs in SLoWPoKES. Dashed lines
    represent the timescales of dynamical dissipation of binaries
    (Weinberg et al.~\cite{Weinberg1987}). The distribution
    shows two populations of wide binaries, suggesting that they were
    either formed differently or have gone through vastly different
    dynamical sculpting.}
\label{Fig: fig1}
\end{figure}

Figure \ref{Fig: fig1} shows the distribution of projected physical
separations and total system mass (inferred from their $r-z$ colors)
for SLoWPoKES systems. There is a distinct bimodality, 
with a break at $\sim$20,000~AU. This is notably the same scale as the
substructure found in star-forming regions and open clusters
(CartWright \& Whitworth~\cite{Cartwright2004}). When compared with dissipation time scales of
wide binaries (Weinberg, Shapiro \& Wasserman~\cite{Weinberg1987}),
the bimodality is suggestive of 
(i) a ``wide'' population that is dynamically stable over $>$10~Gyr
and (ii) an ``ultra-wide'' population of young, loosely-bound 
systems that will dissipate in $\sim$1--2~Gyr
(Dhital et al.~\cite{Dhital2010}). There are two scenarios that explain this
bimodality:

\begin{enumerate}
\item {\it Ultra-wide binaries generally form as hierarchical triples and quadruples in a
  similar manner as other binary systems and with similar separations.} The
  outer orbit then gets wider by transferring angular momentum
  from  inner obit (e.g., Tokovinin~\cite{Tokovinin1997}). Once the orbit is
  larger than a few thousand AU, further energy transfer under the
  influence of Galactic tides, giant molecular clouds, and other stars
  can take place, widening the system to their current observed separations,
  until it eventually dissipates (Weinberg et al.~\cite{Weinberg1987};
  Jiang \& Tremaine \cite{Jiang2010}). A  
  high multiplicity fraction among wide binaries is a requisite in
  this scenario, which further predicts that the widest pairs will be
  most likely to contain a third (or fourth) companion.
\item {\it As suggested by recent numerical simulations, as stars are
    dynamically ejected from their parental cloud core, a small
    fraction will get bound and form ultra-wide systems} (Bate \&
  Bonnell~\cite{Bate2005}; Kouwenhoven et al.~\cite{Kouwenhoven2010};
  Moeckel \& Bate~\cite{Moeckel2010}; Moeckel \& Clarke~\cite{Moeckel2011}). 
  Ultra-wide systems thus formed via dynamical capture would preferentially be
  low-mass but not necessarily have an enhanced hierarchical
  multiplicity. Separations need to be less $\sim$50--80~AU to have
  survived the dynamical ejection from its natal cloud core (Parker \&
  Goodwin~\cite{Parker2010}).
\end{enumerate}

\begin{figure*}
  \includegraphics[angle=0, width=\linewidth]{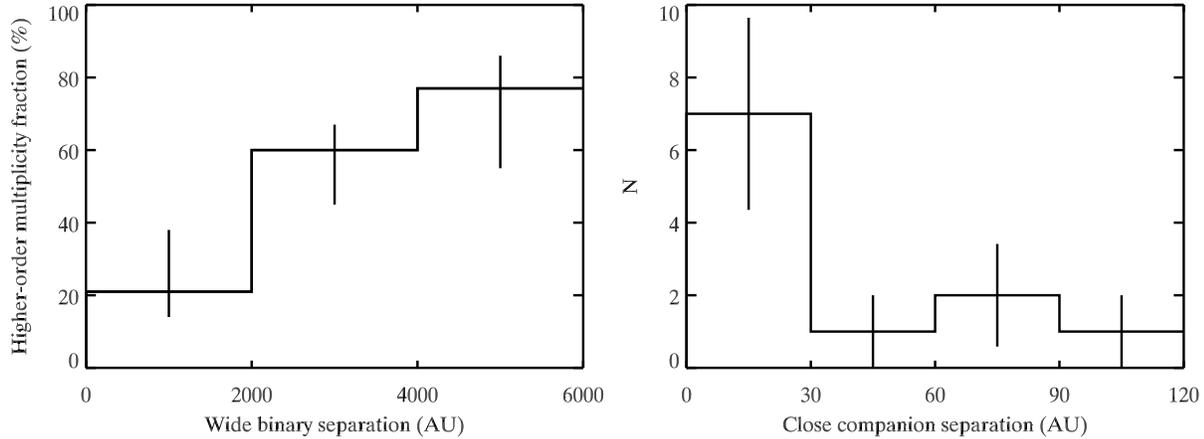}
  \caption{{\it Left:} The higher-order multiplicity fraction
    increases as a function of wide binary separation,
    suggesting that wide binaries form through dynamical widening of
    primordial triples/quadruples.
    {\it Right:} The distribution of physical separations of the close binaries is
    consistent with them having been ejected from clusters and
    getting bound to another ejected star to form a wide binary.
    (Reproduced from Law et al. (\cite{Law2010}).)}
\label{Fig: fig2}
\end{figure*}

Given that we only know the projected
orbital parameters of these very wide systems, discriminating between
even such radically different formation scenarios is not an easy
task. Ensemble properties, preferably ones that are not dependent on
the projection angle, of a statistically significant sample are needed.
Hence, the key discriminant between the two scenarios is the frequency
and arrangement of triples and quadruples among wide binaries.

In a high-resolution imaging study using the Laser Guide Star with
Adaptive Optics at the Keck and Palomar telescopes, we found 45\% of the
observed wide binaries were hierarchical triples and quadruples
(Law et al.~\cite{Law2010}). Moreover, as shown in the left panel of
Figure~\ref{Fig: fig2}, the multiplicity fraction increased from 20\%
at the closest wide binary separations probed ($\sim$1000~AU) to 80\%
at the largest ($\sim$5000 AU), with 2-$\sigma$ significance. For the widest
binaries, the multiplicity fraction is also significantly higher than
the binary fraction observed for isolated low-mass stars in the field
(34--42\%; Fischer \& Marcy~\cite{Fischer1992}; Reid \&
Gizis~\cite{Reid1997}). So while the binaries of 
up to $\sim$1000~AU could have formed via a mechanism similar to other
field stars, wide binaries seem to have formed differently. The
enhanced multiplicity is indicative of a populations formed largely
via dynamically dissipation.

Figure ~\ref{Fig: fig2} (right panel) shows the orbital separations of
the inner binaries in our sample. The distribution peaks sharply at
$<$30~AU; moreover, only one system at physical separation of $>$80
AU, the nominal limit for a binary to survive a dynamical ejection
(Parker \& Goodwin~\cite{Parker2010}). This distribution is consistent with a population
that was formed via dynamical ejection followed by capture.

In conclusion, our data do not rule out either scenario of wide binary
formation but indicate neither mechanism can form all of the observed
wide binaries. A larger sample is needed to confirm our 2-$\sigma$
result of increasing multiplicity with wide binary separation, as well
as to probe systems of lower masses and larger separations.
We conclude that multiple processes, not all of which 
are primordial but are likely to be dynamical in nature, are likely
responsible for the observed distribution of stellar binaries.

\section{A Refined Metallicity Index for M Dwarfs}\label{Sec:metallicity}
While spectral modeling has allowed for metallicity determinations and
well-defined metallicity indices for warmer stars, such efforts in the
late-K and M spectral type regimes (e.g., Hauschildt, Allard \&
Barron~\cite{Hauschildt1999a}) have 
met with notable problems due to the onset of broad molecular lines at
$<$4300 K and due to incomplete molecular line lists. Hence,
the metallicity of low-mass stars remains an elusive parameter to
measure. While near-infrared indices
(Rojas-Ayala et al.~\cite{Rojas-Ayala2010}; Terrien et
al.~\cite{Terrien2012}) will likely lead to accurate 
metallicity estimates, their scope is limited as it is not practical
to acquire near-infrared spectra for large samples of low-mass
stars. However, as the most numerous stellar constituents, M
dwarfs are the best tracers of the formation, chemical, and dynamical
history of the Milky Way. So having an easily observable index tied to
an absolute metallicity scale is imperative. The
$\zeta_{\rm{TiO/CaH}}$ index (L{\'e}pine et al.~\cite{Lepine2007b}) is
ideal for such a purpose as it can be measured from
moderate-resolution optical spectra.

\begin{figure}
  \includegraphics[angle=0, width=\linewidth]{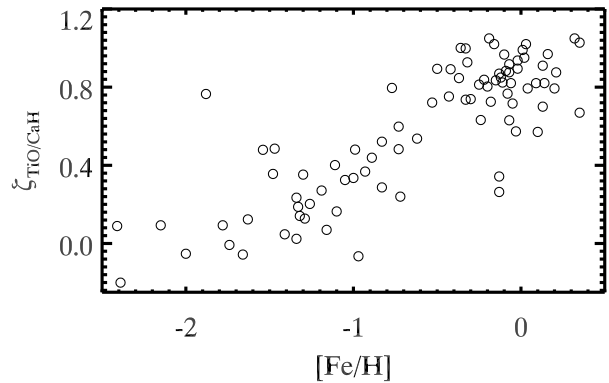}
  \caption{The relative metallicity indicator, $\zeta_{\rm{TiO/CaH}}$
    from Dhital et al.~(\cite{Dhital2012}), vs. the absolute metallicity, [Fe/H],
    from Woolf et al.~(\cite{Woolf2009}). The 
    refined definition of $\zeta_{\rm{TiO/CaH}}$ yields a tighter 
    $\zeta_{\rm{TiO/CaH}}$--[Fe/H] relation but still suffers from
    considerable scatter.}
\label{Fig: fig3}
\end{figure}

From follow up spectra of 113 SLoWPoKES pairs, we found that the
$\zeta_{\rm{TiO/CaH}}$ values for components of a binary are consistent with each other,
within the error bars, for most pairs (Dhital et al.~\cite{Dhital2012}). However, $\zeta_{\rm{TiO/CaH}}$ is
systematically overestimated for higher-mass M dwarfs. Leveraging
the iso-metallicity of components of a pair, we redefined
$\zeta_{\rm{TiO/CaH}}$ by minimizing $\Delta\zeta_{\rm{TiO/CaH}}$. While the change in the
definition is small, it is much more consistent with iso-metallicity
loci that is defined by the observed binaries.

With our new definition of $\zeta_{\rm{TiO/CaH}}$, we also recalibrated the
absolute $\zeta_{\rm{TiO/CaH}}$--[Fe/H] relation from Woolf, L{\'e}pine \&
Wallerstein~(\cite{Woolf2009}), calibrated using a sample of FGK--M
binaries. As Figure~\ref{Fig: fig3}
shows, there is a definitive linear relation between the relative
metallicity indicator, $\zeta_{\rm{TiO/CaH}}$, and the absolute metallicity,
[Fe/H]. However, there remains a considerable scatter. We have
conducted observations of low-metallicity and lower-mass binaries to
further calibrate the $\zeta_{\rm{TiO/CaH}}$--[Fe/H] relation.

\section{A Web-based Data Visualization Portal for Low-mass Dwarfs}\label{Sec:portal}
We have created a new web-based data visualization portal, which is
hosted on a dedicated web server at Vanderbilt
University\footnote{\url{http://slowpokes.vanderbilt.edu}}. The
purpose of this portal is twofold: (i) provide easy access to large
data catalogs of low-mass stars and (ii) allow for easy and fast
visualization of the data without needing to download the entire catalog. We have also
created tools to plot the data on the web browser, to do target
selection for follow-up observations, and to cross-match 
user-inputted objects with the catalogs. Images and web links for the SDSS and
2MASS photometric and, when available, SDSS spectroscopic data are
also included. Currently,
the server hosts four different data sets based on the SDSS survey:
the SLoWPoKES binaries (Dhital et al.~\cite{Dhital2010}), the white dwarf--M dwarf
binary sample (Morgan et al.~\cite{Morgan2012}), and the SDSS spectroscopic
samples of M (West et al.~\cite{West2011}) and L dwarfs (Schmidt et
al.~\cite{Schmidt2010}). This web portal is meant to be a public, live
catalog, so all followup data will be made available as well. 

\section{Future Work}\label{Sec:future}
In Dhital et al.~(\cite{Dhital2010}) we noted that for small angular separations,
proper motions are not required to identify binary systems with a high
level of fidelity. Our Galactic model is able to effectively sift out the
chance alignments to $\sim15\arcsec$. We have thus identified
$\sim$80,000 binary systems with separations of 0.4--15$\arcsec$ and
probability of chance alignment $\leq$5\% (SLoWPoKES-II; Dhital et
al. {\it in preparation}).

The SLoWPoKES catalog has already enabled many programs that use the
pairs as coeval laboratories to explore the fundamental proeprties of
low-mass stars. Ongoing studies include measuring absolute 
metallicity (Dhital et al. \cite{Dhital2012}; Bochanski et al. {\it in preparation}), intrinsic scatter
in activity (Gunning et al. {\it in preparation}), and ages of M dwarfs using
gyrochronology or cooling ages of white dwarfs (Morgan et al. {\it in preparation}). With a
much larger and more diverse SLoWPoKES-II catalog, more follow up
observations should be enabled.

\acknowledgements
SD, AAW, and KGS acknowledge funding support through
NSF grants AST--0909463. AAW also acknowledges support from
AST--1109273. SD would like to thank Cool Stars XVII Organizing
Committee and the NASA Astrobiology Institute for providing funding
for travel to the conference.


\begin{thebibliography}{}

\bibitem[2009]{Abazajian2009}
{Abazajian}, K.\,N. et al.: 2009,  \apjs 182, 543--558

\bibitem[2009]{Bate2009}
{Bate}, M.\,R.: 2005,  \mnras 392, 590--616

\bibitem[2005]{Bate2005}
{Bate}, M.\,R., {Bonnell}, I.\,A.: 2005,  \mnras 356, 1201--1221

\bibitem[2010]{Bochanski2010}
{Bochanski}, J.\,J., {Hawley}, S.\,L., {Covey}, K.\,R., {West}, A.\,A., {Reid},
  I.\,N., {Golimowski}, D.\,A., {Ivezi{\'c}}, {\v Z}.: 2010,  \aj 139,
  2679--2699

\bibitem[2007]{Bochanski2007a}
{Bochanski}, J.\,J., {Munn}, J.\,A., {Hawley}, S.\,L., {West}, A.\,A., {Covey},
  K.\,R., {Schneider}, D.\,P.: 2007,  \aj 134, 2418--2429

\bibitem[2004]{Cartwright2004}
{Cartwright}, A., {Whitworth}, A.\,P.: 2004,  \mnras 348, 589--598

\bibitem[2004]{Chaname2004}
{Chanam{\'e}}, J., {Gould}, A.: 2004,  \apj 601, 289--310

\bibitem[2008]{Connelley2008}
{Connelley}, M.\,S., {Reipurth}, B., {Tokunaga}, A.\,T.: 2008,  \aj 135,
  2496--2525

\bibitem[2010]{Dhital2010}
{Dhital}, S., {West}, A.\,A., {Stassun}, K.\,G., {Bochanski}, J.\,J.: 2010,
  \aj 139, 2566--2586

\bibitem[2012]{Dhital2012}
{Dhital}, S., {West}, A.\,A., {Stassun}, K.\,G., {Bochanski}, J.\,J., {Massey},
  A.\,P., {Bastien}, F.\,A.: 2012,  \aj 143, 67

\bibitem[1991]{Duquennoy1991}
{Duquennoy}, A., {Mayor}, M.: 1991,  A\&A 248, 485--524

\bibitem[1992]{Fischer1992}
{Fischer}, D.\,A., {Marcy}, G.\,W.: 1992,  \apj 396, 178--194

\bibitem[1999]{Hauschildt1999a}
{Hauschildt}, P.\,H., {Allard}, F., {Baron}, E.: 1999,  \apj 512, 377--385

\bibitem[2010]{Jiang2010}
{Jiang}, Y., {Tremaine}, S.: 2010,  \mnras 401, 977--994

\bibitem[2008]{Juric2008}
{Juri{\'c}}, M., et al.: 2008,  \apj 673, 864--914

\bibitem[2010]{Kouwenhoven2010}
{Kouwenhoven}, M.\,B.\,N., {Goodwin}, S.\,P., {Parker}, R.\,J., {Davies},
  M.\,B., {Malmberg}, D., {Kroupa}, P.: 2010,  \mnras 404, 1835--1848

\bibitem[2009a]{Kraus2009a}
{Kraus}, A.\,L., {Hillenbrand}, L.\,A.: 2009a,  \apj 704, 531--547

\bibitem[2009b]{Kraus2009b}
{Kraus}, A.\,L., {Hillenbrand}, L.\,A.: 2009b,  \apj 703, 1511--1530

\bibitem[2010]{Law2010}
{Law}, N.\,M., {Dhital}, S., {Kraus}, A., {Stassun}, K.\,G., {West}, A.\,A.:
  2010,  \apj 720, 1727--1737

\bibitem[2007]{Lepine2007a}
{L{\'e}pine}, S., {Bongiorno}, B.: 2007,  \aj 133, 889--905

\bibitem[2007]{Lepine2007b}
{L{\'e}pine}, S., {Rich}, R.\,M., {Shara}, M.\,M.: 2007,  \apj 669, 1235--1247

\bibitem[2010]{Moeckel2010}
{Moeckel}, N., {Bate}, M.\,R.: 2010,  \mnras 404, 721--737

\bibitem[2011]{Moeckel2011}
{Moeckel}, N., {Clarke}, C.\,J.: 2011,  \mnras 415, 1179--1187

\bibitem[2012]{Morgan2012}
{Morgan}, D.\,P., {West}, A.\,A., {Garc{\'e}s}, A., {Catal{\'a}n}, S.,
  {Dhital}, S., {Fuchs}, M., {Silvestri}, N.\,M.: 2012,  AJ, in press.

\bibitem[2004]{Munn2004}
{Munn}, J.\,A., et al.: 2004,  \aj 127, 3034--3042

\bibitem[2010]{Parker2010}
{Parker}, R.\,J., {Goodwin}, S.\,P.: 2010,  \mnras  (pp.\, 1728--+)

\bibitem[1997]{Reid1997}
{Reid}, I.\,N., {Gizis}, J.\,E.: 1997,  \aj 113, 2246--+

\bibitem[2010]{Rojas-Ayala2010}
{Rojas-Ayala}, B., {Covey}, K.\,R., {Muirhead}, P.\,S., {Lloyd}, J.\,P.: 2010,
  \apjl 720, L113--L118

\bibitem[2010]{Schmidt2010}
{Schmidt}, S.\,J., {West}, A.\,A., {Hawley}, S.\,L., {Pineda}, J.\,S.: 2010,
  \aj 139, 1808--1821

\bibitem[2007]{Stassun2007}
{Stassun}, K.\,G., {Mathieu}, R.\,D., {Valenti}, J.\,A.: 2007,  \apj 664,
  1154--1166

\bibitem[2012]{Terrien2012}
{Terrien}, R.\,C., {Mahadevan}, S., {Bender}, C.\,F., {Deshpande}, R.,
  {Ramsey}, L.\,W., {Bochanski}, J.\,J.: 2012,  \apjl 747, L38

\bibitem[1997]{Tokovinin1997}
{Tokovinin}, A.\,A.: 1997,  Astronomy Letters 23, 727--730

\bibitem[1987]{Weinberg1987}
{Weinberg}, M.\,D., {Shapiro}, S.\,L., {Wasserman}, I.: 1987,  \apj 312,
  367--389

\bibitem[2011]{West2011}
{West}, A.\,A., et al.: 2011,  \aj 141, 97--+

\bibitem[2009]{Woolf2009}
{Woolf}, V.\,M., {L{\'e}pine}, S., {Wallerstein}, G.: 2009,  \pasp 121,
  117--124

\end{thebibliography}
\end{document}